# A Liutex based definition of vortex axis line

Chaoqun Liu[1]*, Yi-sheng Gao[1], Jian-ming Liu[1,2], Yi-fei Yu[1]

1. *Department of Mathematics, University of Texas at Arlington, Arlington, Texas 76019, USA*

2. *School of Mathematics and Statistics, Jiangsu Normal University, Xuzhou 221116, China*

**Abstract:** Eulerian local region-type vortex identification criteria, including the $Q$ criterion, the $\lambda_2$ criterion and the $\lambda_{ci}$ criterion et al., are widely used for vortex identification due to the simplicity in applications. However, most of these criteria are based on a scalar quantity, unable to identify vortex axis (core) lines. On the other hand, the current line-type methods, which seek to extract line-type features such as vortex core lines, are not entirely satisfactory since most of these methods are based on vorticity or pressure minimum which will fail in many cases. To address this issue, a novel Liutex (previously named Rortex) based definition of vortex axis line is proposed in this paper. Mathematically, the vortex axis lines are defined by points where the gradient of Liutex magnitude is aligned with the direction of the Liutex vector, which implies that the cross product of the gradient of Liutex magnitude and the Liutex vector is equal to zero. A preliminary manual process for extracting vortex axis lines is presented. Two test cases, namely Burgers vortex and hairpin vortex, are examined to verify the proposed method.

**Introduction**

Vortices are ubiquitous in nature and can be easily observed from smoke rings to wingtip vortices, from tropical cyclones to even galaxies [1]. In fluid mechanics, it is well recognized that the ubiquity of multi-scale vortical structures, more formally referred to coherent structures [2-4], is one of the most prominent characteristics of turbulent flows and these spatially coherent, temporally evolving structures serve a crucial role in the generation and evolution of turbulence [5-7]. Several different vortical structures, including hairpin vortices [8-10] and quasi-streamwise vortices [3, 11-12], etc., have been identified and intensively studied. Surprisingly, although vortices can be readily observed and intuitively regarded as the rotational/swirling motion of fluids, an unambiguous and generally accepted definition of a vortex has yet to be achieved (and the definition of coherent structures is somewhat ambiguous as well). Since vortices can be considered as elementary structures of organized motion [9] or building blocks for turbulent flows [13], the lack of a rigorous vortex definition is possibly one of the major obstacles to thoroughly understand the mechanism of turbulence generation and sustenance [14].

During the last three decades, a variety of vortex identification methods have been proposed to attempt to answer the deceptively complicated question of vortex definition. An intuitive idea to define a vortex is that closed or spiraling streamlines or pathlines imply the existence of a vortex. One such example is the definition given by Lugt for steady motions [15]: *Any mass of fluid moving around a common axis constitutes a vortex. Mathematically, such motion can be described by closed or spiraling streamlines (or pathlines) if a reference frame exists for which the flow field becomes steady*. Robinson et al. [3] also propose a definition based on instantaneous streamlines which exhibit a roughly circular or spiral pattern. However, these seemingly reasonable definitions will be plagued with a devastating shortcoming: streamlines are not invariant under the Galilean transformation. Another common candidate for vortex definition is the one based on vorticity. The vorticity is well-defined (the curl of the velocity), so a vortex is commonly associated with a vortex filament/tube [16] or a finite volume of vorticity [17]. While vorticity-based methods seem be straightforward, they will run into serious problems in viscous flows, especially in turbulence. The vorticity itself cannot distinguish a real rotational motion from a shear layer and thus the association between regions of strong vorticity and actual vortices can be rather weak in the turbulent boundary layer [18]. In many wall-bounded flows such as the Blasius boundary layer, the magnitude of the vorticity in a near-wall region can be relatively large compared to its surrounding but does not induce any rotational motion. And the maximum vorticity magnitude does not necessarily imply the core of a vortex. Actually, the vorticity magnitude can be considerably decreased along the vorticity lines entering the vortex ring in flat plate boundary layer flows [19]. Furthermore, it is not unusual that the local vorticity vector is not aligned with the direction of vortical structures in turbulent wall-bounded flows, especially in the near-wall regions. Gao et al. [20] point out that the vorticity can be somewhat misaligned with the vortex core direction in turbulent wall-bounded flows. Pirozzoli et al. [21] also show the differences between the local vorticity direction and the vortex core orientation in a supersonic turbulent boundary layer.

The issues of streamline-based and vorticity-based methods for the identification and visualization of vortical structures prompt the development of Eulerian local region-type vortex identification criteria [22]. These local region-type methods are based on the concept that a point is inside a vortex if a criteria is met in that point. Overall, most of the currently popular Eulerian local region-type vortex identification methods are based on the local velocity gradient tensor. More specifically, these criteria are exclusively determined by the eigenvalues or invariants of the velocity gradient tensor and thereby can be classified as eigenvalue-based criteria [23]. For example, in the $Q$ criterion, $Q$ is defined as the residual of the vorticity tensor norm squared subtracted from the strain-rate tensor norm squared, which is equal to the second invariant of the velocity gradient tensor for incompressible flows [24]. The $\Delta$ criterion identifies a vortex as the region where the velocity gradient tensor has complex eigenvalues by the discriminant of the characteristic equation [25]. The $\lambda_{ci}$ criterion can be considered as an extension of the $\Delta$ criterion and uses the (positive) imaginary part of the complex eigenvalue to determine the swirling strength [26]. The $\lambda_2$ criterion defines a vortex as a connected region with two negative eigenvalues of the symmetric tensor of $\mathbf{S}^2 + \mathbf{W}^2$ ( $\mathbf{S}$ and $\mathbf{W}$ represent the symmetric and the antisymmetric parts of the velocity gradient tensor, respectively) [27]. Usually, $\lambda_2$ cannot be expressed in terms of the eigenvalues of the velocity gradient tensor. But when the eigenvectors are orthonormal, $\lambda_2$ can be exclusively determined by the eigenvalues [28]. These methods have several advantages: (1) they are Galilean invariant compared to closed or spiraling streamlines or pathlines; (2) they can discriminate against shear layers, offering more detectable vortical structures; (3) they only depend on the local flow property and can be easy to implement. Nevertheless, the existing eigenvalue-based criteria are not entirely satisfactory. One shortcoming of these criteria is the determination of case-related threshold. It is crucial to determine an appropriate threshold, since different thresholds will present different vortical structures. It has been found that "vortex breakdown" will be exposed with the use of a large threshold for the $\lambda_2$ criterion while will not be observed with a small threshold even if the same DNS data are examined [29]. Due to the sensitivity of threshold change, the educed structures obtained from these criteria should be interpreted with care. To avoid the usage of case-related thresholds, relative values can be employed. One such example is the Omega method proposed by Liu et al [30-31]. The Omega method ($\Omega$) is originated from an idea that the vortex is a region where the vorticity overtakes the deformation and thus $\Omega$ can be defined as a ratio of vorticity tensor norm squared over the sum of vorticity tensor norm squared and deformation tensor norm squared. Accordingly, the Omega method is robust to moderate threshold change and capable to capture both strong and weak vortices simultaneously while most of eigenvalue-based methods are sensitive to threshold change. Another obvious drawback of the existing criteria is the inadequacy of identifying the swirl axis or orientation. Since the existing eigenvalue-based criteria are scalar-valued criteria, only iso-surfaces of a specified threshold can be obtained to detect the vortex while no information about the rotational/swirling axis can be obtained. In addition, eigenvalue-based criteria can be severely contaminated by shearing [23, 32-33]. As pointed out by Gao et al. [23], the existing eigenvalue-based criteria will suffer from this problem as they depend on the imaginary part of the complex eigenvalues.

To address the issues with the existing eigenvalue-based criteria, a novel eigenvector-based Liutex method (previously named Rortex [23, 34] or vortex vector [35]) is proposed. One of the most salient features of Liutex is that Liutex is a systematical definition of the local fluid rotation, including the scalar, vector and tensor interpretations. The scalar version or the magnitude of Liutex represents the local rotational strength (angular velocity). The direction of the Liutex vector, which is determined by the real eigenvector of the velocity gradient tensor, represents the local rotation axis, consistent with the analysis of the instantaneous solution trajectories govern by first-order ordinary differential equations [25-26]. The tensor form of Liutex, rather than the vorticity tensor, represents the real rotational part of the velocity gradient tensor, which can be used for the decomposition of the velocity gradient tensor [36]. Meanwhile, Liutex can eliminate the contamination due to shearing and thus can accurately quantify the local rotational strength, which has been demonstrated for incompressible flows [23]. Moreover, in contrast to the current eigenvalue-based criteria, since Liutex has a vector form to indicate the local rotational axis, Liutex vector field and Liutex lines can be also used to visualize and investigate vortical structures. However, user-specified threshold is still required for the visualization and identification by Liutex iso-surfaces. For this question, a combination of the ideas of the Omega method and Liutex would be a possible solution [37].

Though Eulerian local region-type methods are widely employed for vortex identification due to the simplicity in applications, line-type methods may be preferred to identify vortex core line or vortex skeleton. Several vortex core line detection algorithms are based on the vorticity vector or helicity. Strawn et al. propose an algorithm to collect points that have local maximum vorticity magnitude in a plane normal to the vorticity vector to trace the vortex cores, although the implementation is simplified [38]. Similarly, Banks and Singer develop a two-step predictor-corrector scheme to obtain a series of points that approximate a vortex skeleton [39]. Starting from a seed point, the vorticity direction is first examined to predict the new position of the vortex skeleton and then the candidate location is corrected to the pressure minimum in the plane perpendicular to the vorticity vector. Levy et al. connect points of maximum helicity density to locate vortex core axis [40]. Based on the concept of eigen helicity density, Zhang and Choudhury propose a Galilean invariant scheme to identify vortex tubes and extract vortex core lines in a simulated Richtmyer-Meshkov flow [41]. In addition, pressure minimum is also applied in several line-type methods to identify vortex core lines. Miura and Kida formulate a pseudo-pressure expression to search for sectional local pressure minima and then these points of pressure minima are connected to construct the axial lines [42]. This method is further improved by imposing the additional swirling constraint [43]. Linnick and Rist use $\lambda_2$ to extract vortex core lines by connecting points where has a local minimum in the plane normal to the gradient of $\lambda_2$ [44]. As the real eigenvector corresponding to the real eigenvalue points in the direction of the local swirl axis, Sujudi and Haimes develop an algorithm for identifying the center of swirling flow by finding points where the velocity projected on the plane normal to the eigenvector is zero [45]. Roth analyzes many different line-type methods and introduce a unifying framework for identifying line-type features in terms of a parallel vectors operator [45]. Extensive overview of the currently available vortex identification methods can be found in Refs. [22] and [47].

As already stated, most Eulerian local region-type vortex identification criteria are based on a scalar quantity, unable to detect the vortex axis line which is one of six core elements of the vortex. Meanwhile, these methods are sensitive to threshold change. If the threshold is too small, weak vortices may be captured, but strong vortices will be smeared. If the threshold is too large, weak vortices will be eliminated [13]. This implies that line-type methods would be preferred for vortex identification. On the other hand, the current line-type methods, which aim to extract line-type features such as vortex core lines, are not very successful since most of these methods are based on vorticity or pressure minimum which will fail in many cases. In this paper, a novel Liutex based definition of vortex axis lines is proposed to provide a robust method for identifying vortex axis lines. Mathematically, the vortex axis lines are defined by points where the gradient of Liutex magnitude is aligned with the direction of the Liutex vector, which indicates that the cross product of the gradient of Liutex magnitude and the Liutex vector is equal to zero. A preliminary manual process for extracting vortex axis lines is presented.

The paper is organized as follows. Section 1 is an introduction of Liutex vector and Liutex gradient vector. Section 2 presents the Liutex based definition of vortex axis line and a preliminary manual process for extracting vortex axis lines. In Section 3, two test cases, namely Burgers vortex and hairpin vortex, are examined to verify the proposed method. The conclusions are summarized at the end of the paper.

**1. Liutex vector and Liutex gradient vector**

Liutex is a systematical definition of the local fluid rotation [23, 34, 36, 48]. According to critical point theory [25], if the velocity gradient tensor has complex conjugate eigenvalues, the instantaneous streamline pattern presents a local swirling motion around the direction of the real eigenvector. Therefore, when the velocity gradient tensor $\nabla \vec{v}$ has complex eigenvalues, the direction $\vec{r}$ of the Liutex vector, which represents the local rotation axis, is defined by the real eigenvector of the local velocity gradient tensor and can be written as

$$\nabla \vec{v} \cdot \vec{r} = \lambda_r \vec{r} \quad (1)$$

where $\lambda_r$ is the real eigenvalue. Since the normalized eigenvector is only unique up to a ± sign, a second condition is imposed [48], which reads

$$\vec{\omega} \cdot \vec{r} > 0 \quad (2)$$

where $\vec{\omega}$ represents the vorticity vector. The magnitude of the Liutex vector represents local rotational strength (angular velocity). In the original defition, the magnitude is obtained via somewhat complicated coordinate rotation [34-35]. Recently, Wang et al. [48] introduce a simple, explicit formula to substantially simplify the calculation. This explicit formula for the Liutex magnitude can be expressed as

$$R = \vec{\omega} \cdot \vec{r} - \sqrt{(\vec{\omega} \cdot \vec{r})^2 - 4\lambda_{ci}^2} \quad (3)$$

Here $\lambda_{ci}$ is the imaginary part of the complex eigenvalue of the velocity gradient tensor. Thus, the Liutex vector $\vec{R}$ is obtained by

$$\vec{R} = R\vec{r} \quad (4)$$

Intuitively, it is expected that the vortex axis line should consist of some kind of local extreme points located on the plane perpendicular to the vortex core line. Since the gradient of a physical quantity is commonly used to find the local extreme point and the Liutex magnitude can represent the accurate local rotational strength, it is natural to choose the gradient of the Liutex magnitude for identifying vortex axis lines.

**Definition 1.** The Liutex gradient vector is defined by the gradient of the Liutex magnitude, which reads

$$\nabla R = \begin{bmatrix} \dfrac{\partial R}{\partial x} \\ \dfrac{\partial R}{\partial y} \\ \dfrac{\partial R}{\partial z} \end{bmatrix} \tag{5}$$

**2. The Liutex based definition of vortex axis line**

The straightforward integration of the Liutex gradient vector will always result in massive discontinuous segments rather than distinct and continuous vortex axis lines. Actually, the usage of the gradient of other common scalar quantities such as $Q$, $\lambda_{ci}$, $\lambda_2$ or even $\Omega$ will also fail owing to the ubiquity of stationary points (zero gradient) in the whole flow fields. To seek local extreme points, another possible candidate can be derived from the concept of parallel vectors operator [46], which can be given by

$$S = \{\vec{x} : (\boldsymbol{H} \cdot \nabla F) \times \nabla F = 0 \,\&\, \eta_2 > 0\} \tag{6}$$

Here, the set $S$ consists of points $\vec{x}$ satisfying Eq. (6). $F$ denotes a scalar quantity like $\lambda_2$ used in Ref. [44], $\boldsymbol{H}$ the Hsssian matrix of $F$ and $\eta_2$ the second eigenvalue of $\boldsymbol{H}$ (ascending order). Eq. (6) indicates the vectors $\boldsymbol{H} \cdot \nabla F$ and $\nabla F$ are parallel to each other. It is easy to show that $\nabla F$ is the eigenvector of the Hsssian matrix $\boldsymbol{H}$ and the points satisfying Eq. (6) are local extreme on the plane perpendicular to the direction of $\nabla F$. But it has been pointed out in Ref. [46] that the solutions of Eq. (6) are actually loci of zero curvature. Worse still, the numerical noise will dramatically undermine the accuracy even if high order methods are applied for the calculation of the Hessian matrix [44], so makes this definition impractical.

Fortunately, the definition of Liutex not only contains the magnitude, but also the direction. The Liutex lines are continuous inside the vortex region (with zero threshold) and have been successfully used to demonstrate the skeleton of hairpin vortex [23, 34-35]. Hence, we can combine the Liutex lines and the Liutex gradient lines to extract vortex core lines. Based on the mathematical property of the Liutex gradient vector and the unique definition of the Liutex vector, we can give the following definition:

**Definition 2.** The vortex core line is defined as the points of $S$ which satisfy the condition

$$S = \{\vec{x} : \nabla R \times \vec{r} = 0 \ \& \ \vec{r} > 0\} \tag{7}$$

where $\vec{r}$ represents the direction of the Liutex vector, i.e. the (normalized) real eigenvector. (It should be noted that every point in the pure rigid body rotation satisfies Eq. (7), so no unique vortex axis line will be obtained. But this type of fluid motion occurs infrequently in real flows.) Eq. (7) implies the vectors $\nabla R$ is aligned with the direction of the Liutex vector $\vec{r}$. Theoretically, the vortex axis lines can be obtained by connecting all points satisfying Eq. (7). However, the computational error made it difficult to solve Eq. (7) directly. In practice, we can find a start point in the vortex which satisfies Eq. (7) and then integrate the Liutex vector passing through this point to obtain a Liutex line which is a very close approximation to the vortex core line. Based on the observation that all the Liutex gradient lines will converge to a gathering line which can be considered as a vortex axis line, the start point can be determined by (approximately) picking up a point on the gathering line. For two-dimensional flows, the vortex axis line degrades a vortex core point and the rotation axis always points out of the 2D plane. In this case, Eq. (7) becomes

$$S_{2D} = \{\vec{x} : \nabla R = 0\} \tag{8}$$

which is also a reasonable definition. Thus, Definition 2 is a universal definition of vortex axis line.

As a result, a preliminary manual process for extracting vortex axis lines is presented here:

Step (1). Draw iso-surfaces by the Omega method or the Liutex magnitude;

Step (2). On the iso-surface, find the gathering line of the Liutex gradient lines;

Step (3). Find a point on the gathering lines and draw the Liutex line passing through this point.

As will be shown in the following section, despite of manual process, the proposed method is quite robust and can avoid any user-specified threshold. The automatic algorithm for identifying vortex axis lines is left for future study.

## 3. Test cases

First, the Burger vortex is examined. The Burger vortex is an exact steady solution of the Navier–Stokes equation and has been widely applied for modelling fine scales of turbulence. The velocity components in the cylindrical coordinates for a Burger vortex are given by

$$v_r = -\xi r \tag{9}$$

$$v_\theta = \frac{\Gamma}{2\pi r}\left(1 - e^{\frac{-r^2 \xi}{2\nu}}\right) \tag{10}$$

$$v_z = 2\xi z \tag{11}$$

where $\Gamma$ represents the circulation, $\xi$ the axisymmetric strain rate, and $\nu$ the kinematic viscosity. The Reynolds number can be defined as

$$\mathrm{Re} = \frac{\Gamma}{2\pi\nu} \tag{12}$$

Hence, the velocity in the Cartesian coordinate system will be written as

$$u = -\xi x - \frac{\Gamma}{2\pi r^2}\left(1 - e^{\frac{-r^2\xi}{2\nu}}\right) y \tag{13}$$

$$v = -\xi y + \frac{\Gamma}{2\pi r^2}\left(1 - e^{\frac{-r^2\xi}{2\nu}}\right) x \tag{14}$$

$$w = 2\xi z \tag{15}$$

Accordingly, the analytical expressions of the Liutex magnitude can be easily obtained as

$$R = 2\,\mathrm{Re}\,\xi\zeta \tag{16}$$

where

$$\tilde{r} = r\sqrt{\xi/\nu} \tag{17}$$

$$\zeta = \frac{1}{\tilde{r}^2}\left[(1+\tilde{r}^2)e^{-\frac{\tilde{r}^2}{2}} - 1\right] \tag{18}$$

For the existence of Liutex, $\zeta$ should be larger than zero, which yields a non-dimensional vortex radius size of $\tilde{r}_0 = 1.5852$, consistent with the result of Ref. 28. And the direction of the Liutex vector always points to the positive z direction for any point inside the vortex. According to Eqs. (16) and (18), the Liutex magnitude has a maximum along the z-axis and the Liutex gradient vector is perpendicular to the Liutex vector elsewhere inside the vortex. Therefore, there exists only one vortex axis line along the z-axis, consistent with the condition imposed by Eq. (7), which means the proposed definition of vortex axis line is valid for the Burgers vortex. Figure 1 shows the vortex axis line and several Liutex gradient lines on the xy-plane.

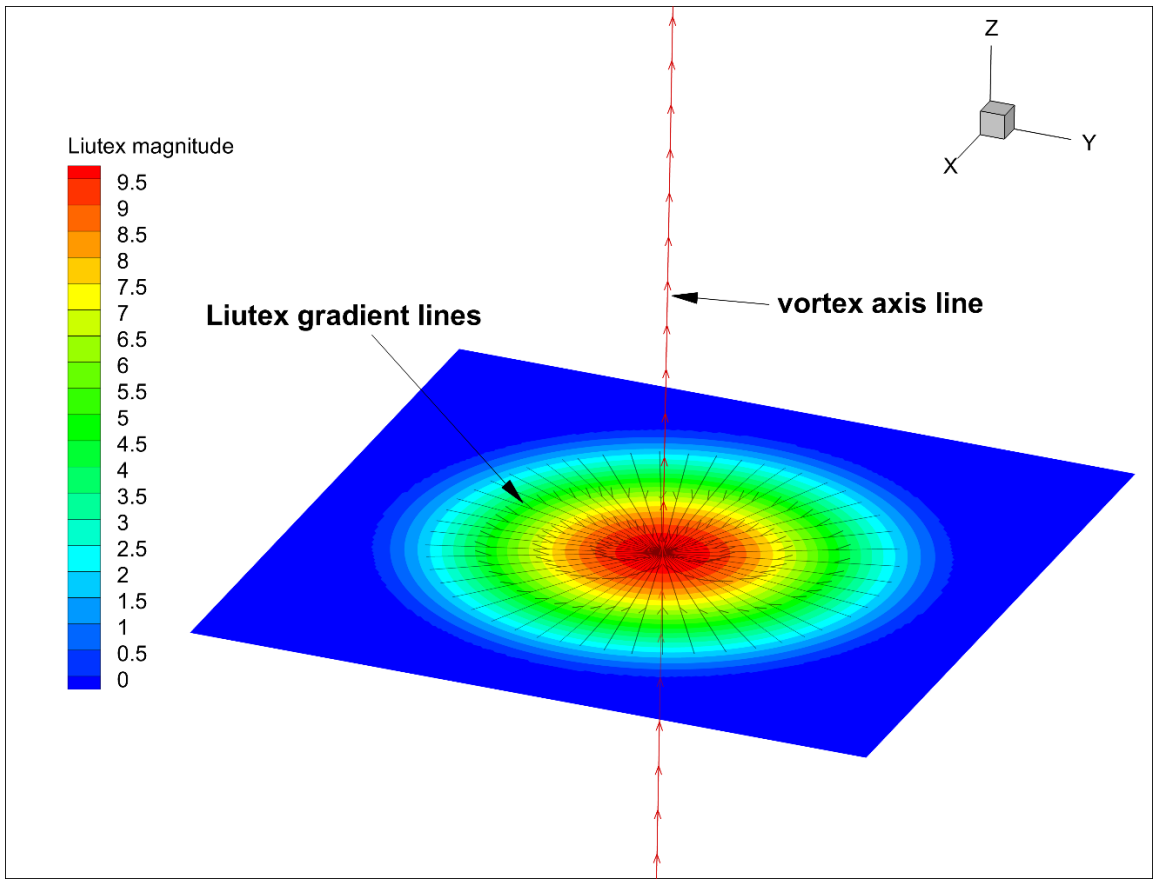

Fig. 1 Vortex axis line and Liutex gradient lines for the Burgers vortex

In the following, the proposed method is applied to extract vortex axis lines for hairpin vortices. The data is obtained from direct numerical simulation (DNS) of the late transition of flat plate boundary layer [5]. The simulation is performed with about 60 million grid points and over 400 000 time steps at a freestream Mach number of 0.5. For the detailed computational setting, one can refer to Ref. [5].

At the first step, the Liutex gradient lines can be obtained from the stream tracer in the visualization software (see Figures 2 and 3(a)). It can be found that the Liutex gradient lines pass through the local extrema on the slice. Hence, the vortex core line can be determined by the Liutex gradient line. However, the gradient of the Liutex magnitude would be zero on the vortex core line, where the point could be the exact three-dimensional extrema. So, the Liutex gradient line will break down on the vortex core line (see Figure 2). But the gathering line of the Liutex gradient lines can help us to find the points on the vortex core. With these points, we can easily get the Liutex line, which is the vortex core line (see Figure 3(b)). Figure 3(c) shows the Liutex gradient line is completely aligned with the Liutex line. In Figure 3(d), we can find the Liutex vortex strength can change along the vortex axis line. Figure 4 shows the vortex core lines and the two-dimensional streamlines, indicating that the streamlines rotate around the vortex axis line.

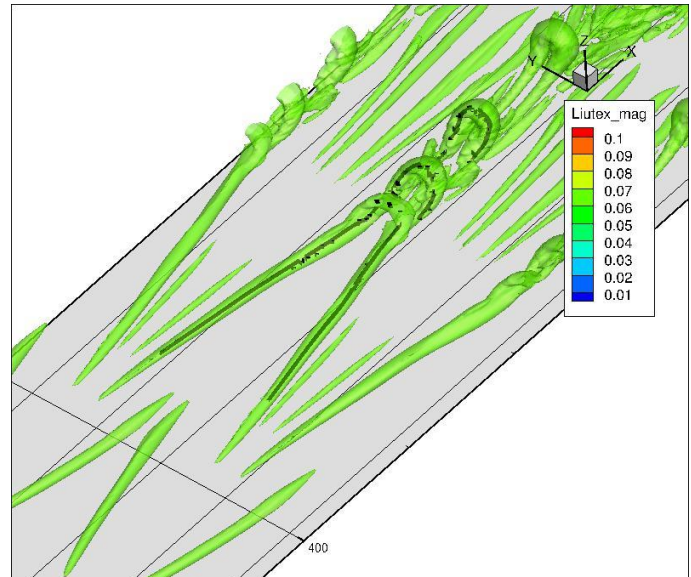

Fig. 2 Liutex gradient lines for hairpin vortices

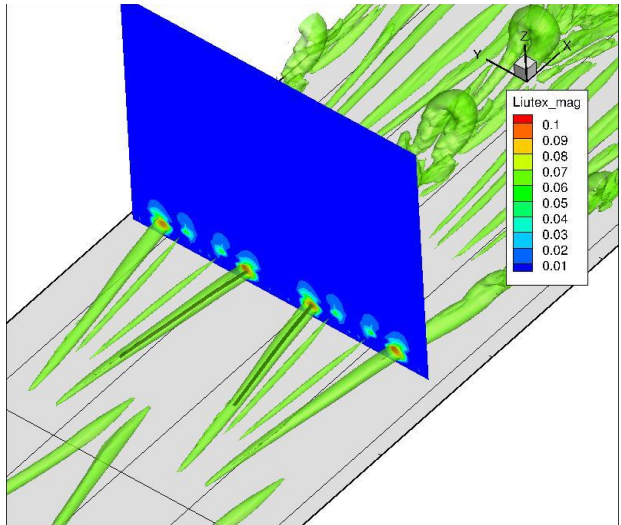

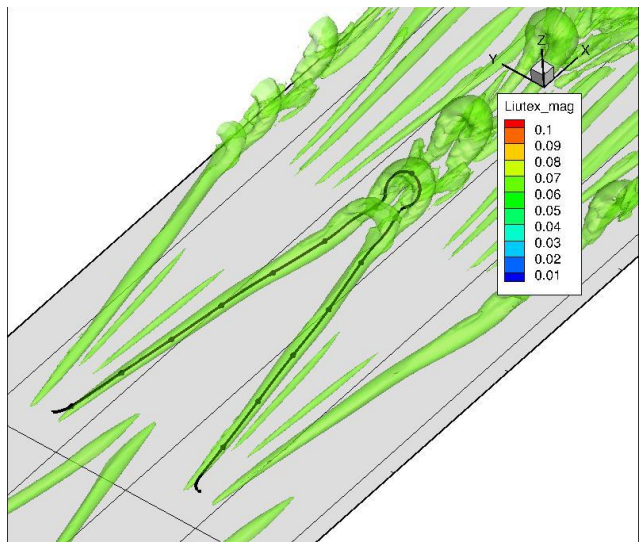

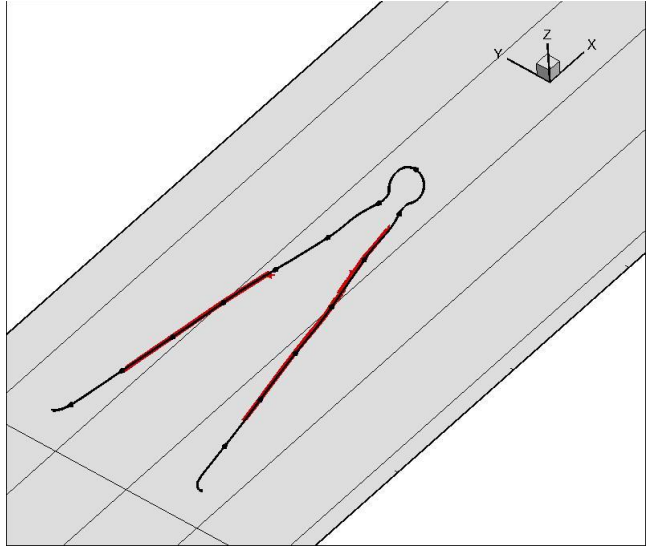

(c)

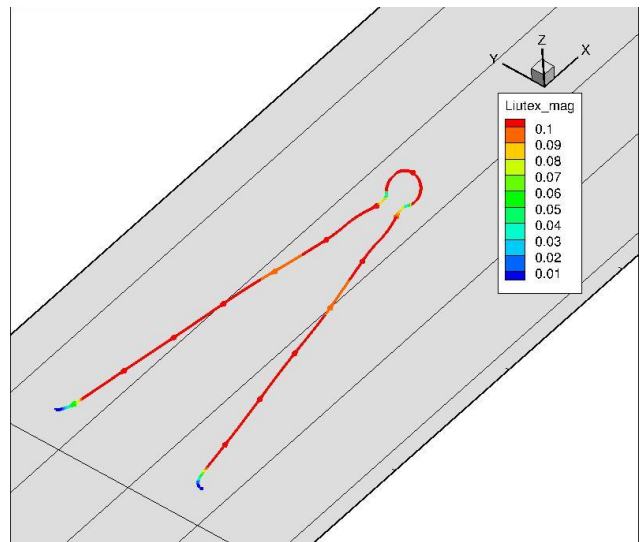

(d)

Fig. 3 (a) Liutex gradient line; (b) vortex core line; (c) Liutex gradient line (red) and vortex core line (black); (d) The magnitude of Liutex on vortex core line

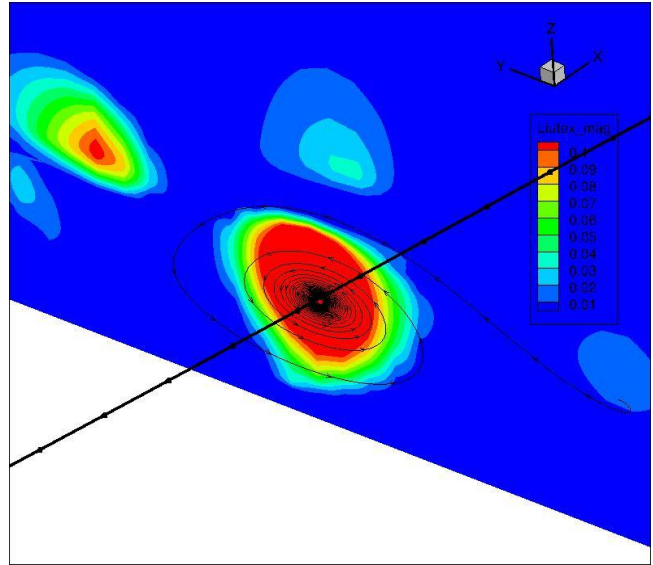

(a)

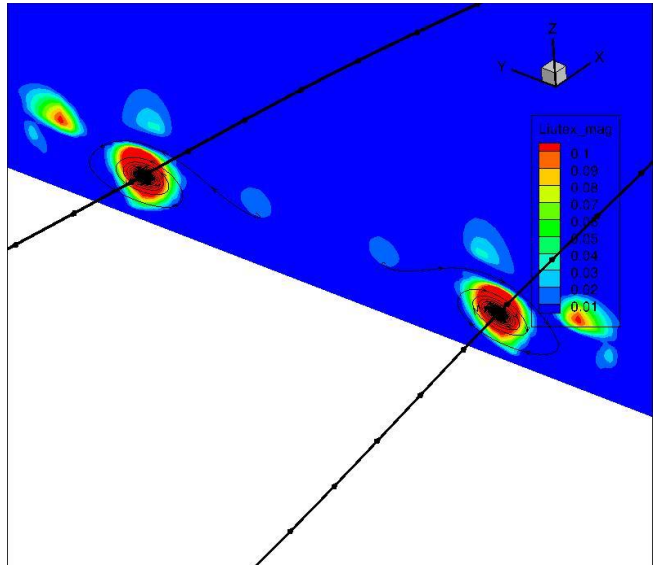

(b)

Fig. 4 Vortex core lines and 2D streamlines on the slice

## 4. Conclusion

In this paper, a Liutex based defition of vortex axis line is presented, which requires that the Liutex gradient vector is aligned with the Liutex vector. Based on the observation that all the Liutex gradient lines will converge to a gathering line which can be considered as a vortex axis line, a preliminary manual process for extracting vortex axis lines is introduced and verified by the Burgers vortex and hairpin vortices. The results demonstrate that the present method can accurately extract vortex core lines without any user-specified threshold.

**Acknowledgement**

This work is supported by the Department of Mathematics at University of Texas at Arlington and AFOSR grant MURI FA9559-16-1-0364. Dr Jianming Liu is partly supported by the National Natural Science Foundation of China (Grant No. 91530325) and the Natural Science Foundation of the Jiangsu Higher Education Institutions of China (Grant No.18KJA110001) and the Visiting Scholar Scholarship of the China Scholarship Council (Grant No. 201808320079). The authors are grateful to Texas Advanced Computational Center (TACC) for providing computation hours. This work is accomplished by using code DNSUTA developed by Dr. Chaoqun Liu at the University of Texas at Arlington.